\documentstyle[11pt]{article}

\begin{document}

\thispagestyle{empty}

\begin{center}
{\Large {\bf Simple approximation for the starting-energy-independent
two-body effective interaction with applications to $^6\mbox{Li}$}}

\vspace{0.2in}

D. C. Zheng and B. R. Barrett\\
\begin{small}
{\it Department of Physics, University of Arizona, Tucson, AZ 85721}
\end{small}

\vspace{0.1in}

J. P. Vary \\
\begin{small}
{\it Department of Physics and Astronomy, Iowa State University, Ames,
					IA 50011}
\end{small}

\vspace{0.1in}

R. J. McCarthy \\
\begin{small}
{\it Department of Physics, Kent State University, Ashtabula,
Ohio 44004}
\end{small}

\end{center}

\begin{abstract}
We apply the Lee-Suzuki iteration method to calculate
the linked-folded diagram series for a new Nijmegen local NN
potential. We obtain an exact starting-energy-independent
effective two-body interaction for a multi-shell, no-core,
harmonic-oscillator model space.
It is found that the resulting effective-interaction matrix elements
can be well approximated by the Brueckner G-matrix elements evaluated
at starting energies selected in a simple way.
These starting energies are closely
related to the energies of the initial two-particle states
in the ladder diagrams. The ``exact'' and approximate effective interactions
are used to calculate the energy spectrum of $^6\mbox{Li}$ in order to test
the utility of the approximate form.
\end{abstract}

\pagebreak

\section{Introduction}
Conventional shell-model (SM) calculations often assume an inert core
with a few valence nucleons as active particles. The calculation of the
two-body effective interaction for the valence nucleons can be conveniently
divided into three steps: (1) Calculate the Brueckner reaction
matrix G \cite{bruc} from a realistic NN potential;
(2) Calculate the two-body Q-box \cite{qbox} from the G-matrix;
and (3) Calculate the folded diagrams \cite{brandow} from the Q-box.
The second step provides major difficulties, because one is unable to
evaluate the core-polarization diagrams to all orders and
there is no sign of convergence within the lowest few orders \cite{barrett}.
This difficulty might be avoided by the use of a no-core model space
\cite{nocore},
for which all the nucleons in a nucleus are treated as active.
In such a model space, because there are no hole lines,
all the core-polarization diagrams are absent and the two-body
Q-box reduces to the G-matrix. The folded diagrams can be calculated from
the Q-box by using iteration methods proposed by Kuo and Krenciglowa \cite{kk}
or by Lee and Suzuki \cite{ls}.
Beyond these issues are the largely unexplored questions on the
role of effective many-body forces in no-core model spaces.

In previous works \cite{jaqua,zheng}, we calculated the low-lying
energy spectra for a few light nuclei by employing no-core model
spaces. We approximated the two-body effective interaction by
the G-matrix and neglected the folded diagrams.
Because of this approximation, our calculations involved
the starting energy as a parameter,
which, in one study \cite{zheng},
was chosen to fit the nuclear binding energy.

In this work, we will calculate the G-matrix
using an improved version of the Nijmegen potential (NijmII) \cite{nijm}
and we follow the Lee and Suzuki method
\cite{ls} to sum the two-body folded diagrams to all orders.
The resulting starting-energy-independent
two-body effective interaction is not Hermitian but its
non-Hermiticity is found to be extremely small.
We obtain the Hermitian effective interaction
$v_{\rm eff}^{(2)}$ to be used in SM calculation
by taking the average of the non-Hermitian effective interaction and
its conjugate. It is pointed out in Ref.\cite{kuopl} that this is
an excellent approximation.

We will also discuss the choice of
the two-nucleon Hamiltonian $H^{(2)}$ employed in the
G-matrix calculation, which
determines the intermediate-energy spectrum in the ladder diagrams.
Currently, due to uncertainties in the
optimal choices of one-body potentials and methods for treating
the spurious center-of-mass motion,
there is no generally accepted $H^{(2)}$.
It is obvious that the two-body effective interactions $v^{(2)}_{\rm eff}$
depend on $H^{(2)}$, so that one wishes to employ the
$H^{(2)}$ which best represents the physics of the
two-nucleon subsystem in the nuclear medium.
We will present a choice which is physically motivated yet retains
simplicity for calculations.

We will furthermore
introduce an approximation scheme which allows us to obtain easily,
an effective two-body interaction for a no-core model space directly from the
starting-energy-dependent G-matrix without evaluating
the folded diagrams. We will demonstrate that the resulting approximate form
is an improvement over the procedure we introduced in
Ref.\cite{zheng} and more closely represents the exact theoretical
$v^{(2)}_{\rm eff}$.

\section{G-matrix and Two-body Effective Interaction}
\label{gv}
Assuming that there are only two-body interactions among nucleons
in a nucleus, the nuclear Hamiltonian for an $A$-nucleon nucleus
can be written as:
\begin{equation}
H = \left(\sum_{i=1}^A t_i -T_{\rm CM}\right)
	+ \sum_{i<j}^A v_{ij} ,   		\label{ha}
\end{equation}
where $t$ is the one-body kinetic energy, $T_{\rm CM}$
is the center-of-mass (CM) kinetic energy of the nucleus, $v$ is
the two-body NN potential.

The expression for the Brueckner reaction matrix G \cite{bruc}
can be generally written in the following form:
\begin{equation}
G_{12}(\omega) = v_{12} + v_{12}\frac{Q}{\omega-H^{(2)}}v_{12},
				\label{gg}
\end{equation}
where $Q$ is the Pauli projection operator and $\omega$ is the starting
energy. The two-nucleon Hamiltonian $H^{(2)}$
represents the dynamics of the two-particle subsystem
in the nuclear medium generated by the remaining
($A$-2) nucleons (also referred to as spectators).
It makes sense to optimize the description of this two-particle
subsystem in order to minimize the effects of the many-body
effective interactions which we plan to neglect.
Clearly the role of the medium on the single-particle states
needs to be included and one is, therefore, led to introduce
one-body potentials into $H^{(2)}$ yielding:
\begin{equation}
H^{(2)} = (t_1+t_2) + v_{12} + {\cal V}_1 + {\cal V}_2
       \equiv (h_1+h_2) + v_{12},  \label{h2V}
\end{equation}
where $h=t+{\cal V}$ is the single-particle (SP) Hamiltonian
with $t$ the one-body kinetic energy and
${\cal V}$ the mean field generated by the spectators.

A physically motivated ${\cal V}(\mbox{\boldmath $r$})$
could be the Hartree-Fock (HF) self-consistent mean field or
the phenomenological Woods-Saxon (WS) well with appropriate depth,
width and surface thickness.
However, to employ either the HF field or the WS well is
a very computationally demanding project for a realistic force
$v_{12}$. For convenience, we assume that
the mean field ${\cal V}$ can be approximated
by a {\it shifted} harmonic-oscillator (HO) potential
(see Fig.1), namely,
\begin{equation}
{\cal V}(\mbox{\boldmath $r$}) \simeq {\cal V}_{\rm HO}^{\rm shifted}(r)
\equiv -V_0 + \frac{1}{2}m\Omega^2 r^2 = -V_0 + u^{\rm HO}(r).   \label{appr}
\end{equation}
The use of the HO potential in conjunction of the HO basis
not only simplifies the G-matrix calculation
\cite{bhm} but also facilitates the treatment of the spurious center-of-mass
motion.

Note that, as far as the low-lying states are concerned,
one need not be greatly concerned with the obvious fact that,
when $r$ goes to infinity,
${\cal V}(\mbox{\boldmath $r$})$ vanishes while
$u^{\rm HO}(r)$ becomes infinite.
Actually, in the low-lying states, the nucleons in a nucleus remain
primarily within the nuclear radius $R_A$,
so the shape of ${\cal V}(r)$ for
large $r$, say, $r>2R_A$, plays a less significant role in the
bound-state spectrum.

With the approximation stated in Eq.(\ref{appr}),
the two-nucleon Hamiltonian (\ref{h2V}) becomes
\begin{equation}
H^{(2)} \simeq (t_1+t_2) + v_{12} + (u^{\rm HO}_1 -V_0) + (u^{\rm HO}_2 -V_0)
       \equiv (h^{\rm HO}_1+h^{\rm HO}_2) + v_{12} - 2V_0,
\end{equation}
where $h^{\rm HO} = t+u^{\rm HO}$ is the pure HO SP
Hamiltonian. The corresponding G-matrix (\ref{gg}) becomes
\begin{equation}
G_{12}(\omega) \simeq v_{12} + v_{12} \frac{Q}{\omega'
	- (h^{\rm HO}_1+h^{\rm HO}_2+v_{12})} v_{12}
	\equiv G^{\rm HO}_{12}(\omega'),   \label{go}
\end{equation}
where $\omega'\equiv \omega+2V_0$. In writing the above equation,
we have added a constant ($2V_0$) to both the starting energy $\omega$
and the two-nucleon Hamiltonian $H^{(2)}$. Obviously this does not change
the result for the G-matrix which depends only on the difference
between $\omega$ and $H^{(2)}$.

For no-core model spaces,
the starting-energy-independent
two-body effective interaction $v_{\rm eff}^{(2)}$ is the ladder
diagram series (G-matrix) plus the folded diagrams.
If we approximate the G-matrix
by $G^{\rm HO}(\omega')$ as in Eq.(\ref{go}),
$v_{\rm eff}^{(2)}$ is written as
\begin{equation}
v_{\rm eff}^{(2)} \simeq G^{\rm HO}(\omega')
	+ ({\rm Folded \hspace{0.05in} diagrams}). \label{fold}
\end{equation}
The folded diagrams can be evaluated by employing the
iteration methods proposed in Refs.\cite{kk,ls}.
We will use the ``vertex-renormalization'' procedure of Ref.\cite{ls}.
The input to this method consists of the
$G(\omega')$ and its derivatives with respective to $\omega'$,
whose values are taken at an
arbitrary (in principle) but fixed starting energy $\omega'$.

We use a HO SP basis with $\hbar\Omega$=18 MeV and
a no-core model space containing the first 4 major shells ($0s$, $0p$,
$1s$-$0d$ and $1p$-$0f$). The $Q$ operator in Eq.(\ref{go}) is defined
to forbid the scattering of the two particles
into an intermediate state inside the model space (i.e., $Q$=0).

For $v_{12}$,
we adopt a new Nijmegen local NN potential (NijmII) \cite{nijm},
which was fitted to the world NN scattering data
with a nearly optimal $\chi^2$ per degree of freedom (1.03 per datum).
Other potentials (Reid93 and AV18)
obtained by fitting the same data have
a comparable $\chi^2$ and yield similar deuteron and triton properties
(see Ref.\cite{friar} for more details).
The derivatives of the G-matrix are calculated
numerically through ninth order using 11 sets of $G(\omega')$
with $\omega'$ ranging from -75 MeV to +75 MeV,
in steps of 15 MeV. In Table 1 under column ``$v^{(2)}_{\rm eff}$'',
we list a few diagonal two-body matrix elements (TBMEs) of the resulting
effective interaction. For comparison and further discussion below,
we also present, in Table 1, G-matrix elements at selected
values of $\omega'$, and at a state-dependent choice of $\omega'$.

\begin{table}
\begin{small}
\caption{Selected diagonal TBMEs for the
starting-energy-independent two-body effective interaction
$v^{(2)}_{\rm eff}$.
These matrix elements should be compared with the G-matrix elements
listed in the table for three different starting energies.
Values in italics indicate values in good agreement with $v^{(2)}_{\rm eff}$.
Note that $2\epsilon^{\rm HO}_{0s}=3\hbar\Omega =54$ MeV,
$2\epsilon^{\rm HO}_{0p}=5\hbar\Omega =90$ MeV,
$2\epsilon^{\rm HO}_{0d}=7\hbar\Omega =126$ MeV.}
\begin{center}
\begin{tabular}{c|c|cccc}\hline\hline
State & $v_{\rm eff}^{(2)}$ & $G(\omega'=30)$
	& $G(\omega'=75)$ & $G(\omega'=110)$ & ``Approx.'' \\ \hline
$(0s_{1/2}^2)^{J=0,T=1}$ & -8.75&{\it -8.73} &-9.05 &-9.47 & -8.75 \\
$(0s_{1/2}^2)^{J=1,T=0}$ &-11.78&{\it -11.70}&-14.18&-20.64&-11.83 \\
$(0p_{3/2}^2)^{J=0,T=1}$ & -3.87 &-3.54 &{\it -3.86} &-4.35& -3.81 \\
$(0p_{3/2}^2)^{J=1,T=0}$ & -2.28 &-1.33 &{\it -2.34} &-4.23& -2.17 \\
$(0d_{5/2}^2)^{J=0,T=1}$ & -1.79 &-1.09 &-1.38 &{\it -1.68}& -1.62 \\
$(0d_{5/2}^2)^{J=1,T=0}$ & -0.78 & 0.76 & 0.05 &{\it -0.84}& -0.65 \\
							\hline\hline
\end{tabular}
\end{center}
\end{small}
\end{table}

\section{Approximation to $v_{\rm eff}^{(2)}$}

As shown in Eq.(\ref{fold}), for a {\it no-core} model space,
the $\omega$-independent
$v_{\rm eff}^{(2)}$ is the sum of the ladder diagrams and the
folded diagrams, both of which depend separately  on the starting energy
$\omega$ (or $\omega'$). When the folded diagrams are ignored, as
was often done in the past, one approximates
$v_{\rm eff}^{(2)}$ by $G(\omega')$. It is obvious that
the contribution of the folded diagrams correlates
with $\omega$. Below, we show that a particular choice of $\omega$
minimizes the root-mean-square (rms)
contributions of the folded diagrams to the TBMEs
of $v_{\rm eff}^{(2)}$. We further
show that a state-dependent choice of $\omega'$
yields a remarkably good approximation to $v_{\rm eff}^{(2)}$.

In the full theory of the effective
Hamiltonian, one has, in principle, independence of the mean field
${\cal V}$ and of $\omega$. However, in practical calculations, to
minimize the need to calculate higher-order processes, one
wishes to make physically sensible choices for
these quantities such as discussed above in the case of ${\cal V}$.

We now consider arguments that may be presented to suggest
a phenomenological choice for $\omega$ that could also
simplify the calculation of the effective two-body interaction.
It is generally accepted that the starting energy $\omega$
represents the initial energy $E_2$ of the two nucleons
in the nuclear medium. In G-matrix calculations, the two nucleons
are treated as two interacting particles
moving in the mean field ${\cal V}$. We can think of
the energy $E_2$ for a two-particle state, which is
predominantly $|ab\rangle_{J,T}$ ($a$ and $b$ are the HO
SP orbitals), as given by:
\begin{equation}
E_2 = \epsilon_a+\epsilon_b+\Delta,   \label{e2}
\end{equation}
where $\epsilon_a$ and $\epsilon_b$ are eigenenergies of the
one-body Hamiltonian $(t+{\cal V})$. The quantity $\Delta$
represents the interaction energy and
depends implicitly on the two-particle state $|ab\rangle_{J,T}$.

When the mean field ${\cal V}$ is approximated by the shifted
HO potential as we did in Eq.(\ref{appr}), Eq.(\ref{e2}) becomes:
\begin{equation}
E_2 \simeq (\epsilon^{\rm HO}_a-V_0)+(\epsilon^{\rm HO}_b-V_0)
			+\Delta,   \label{e2ho}
\end{equation}
where $\epsilon^{\rm HO}_a$ and $\epsilon^{\rm HO}_b$ are the HO SP
energies [$\epsilon^{\rm HO}_i = (2n_i+l_i+\frac{3}{2})\hbar\Omega$
with $i=a,b$].

Since the shifted starting energy $\omega'$ used in
Eq.(\ref{go}) for the G-matrix
is related to the original starting energy $\omega$ through
$\omega'=\omega+2V_0$, we have the following equation for
$\omega'$:
\begin{equation}
\omega' \simeq \epsilon^{\rm HO}_a+\epsilon^{\rm HO}_b
			+\Delta.   \label{wp}
\end{equation}
Note that although $E_2$ and, thus, $\omega$ are
negative for a two-particle state bound in the nucleus,
$\omega'$ is not necessarily
negative, because the quantity $V_0$ (approximately representing
the depth of the potential well
generated by the spectators) is always positive.
This provides a partial justification to
the choice of $\omega'$ made in Ref.\cite{zheng}, where
G-matrices at a positive $\omega'$ are found to yield
the approximately correct nuclear binding energies.

We, therefore, expect the G-matrix for the starting
energy $\omega'$ given by Eq.(\ref{wp}) to be a reasonable
approximation to $v^{(2)}_{\rm eff}$.
This is clearly demonstrated in Table 1, where the
TBMEs of $G(\omega')$ for three values of $\omega'$ are listed.
{}From Table 1, one can see that
the listed TBMEs of $v^{(2)}_{\rm eff}$
can be approximated by those of $G^{\rm HO}(\omega')$
at a starting energy $\omega'$ given by Eq.(\ref{wp}) with
$\Delta$ ranging from --25 MeV to --15 MeV (numbers in italic).
As we mentioned before,
$\Delta$ more or less represents the state-dependent
contribution to the two-nucleon energy
from the NN interaction.

If one picks $\Delta$ for each matrix element such that
$G(\omega')=V^{(2)}_{\rm eff}$, then from
Table 1, one can also see that
$\Delta$ is generally larger in magnitude for the $J$=1,
$T$=0 channel than for the
$J$=0, $T$=1 channel. This is physically sensible since the NN interaction
for the former channel is more attractive.
It is also obvious from the table that $\Delta$ is larger
in magnitude for lower-lying two-nucleon states (e.g., $0s^2_{1/2}$)
than for higher-lying ones (e.g., $0p^2_{3/2}$ or $0d^2_{5/2}$),
which accounts for the fact that the NN interaction is stronger
for the former states.

The observations above lead us to suggest a simple state-dependent
choice for $\omega'$. That is, we suggest using Eq.(\ref{wp})
with $\Delta$ taken as a single constant for all states for
simplicity. In this way, we hope that $G(\omega')$
will become a good approximation to $v^{(2)}_{\rm eff}$.
We see that this is, indeed, the case by comparing the ``exact''
results in the first column of Table 1 with the ``approximate''
results [i.e. those obtained with $\omega'$ of Eq.(\ref{wp})
using $\Delta=-21$MeV] in the last column of Table 1.

To further illustrate the difference between a state-independent and our
state-dependent choice of $\omega'$, we define
an rms deviation of the matrix elements of
$G$ from those of the starting-energy-independent $v_{\rm eff}^{(2)}$ as
\begin{equation}
\delta(\omega') = \left\{ \frac{1}{N}
	\sum \left[ \langle ab|v_{\rm eff}^{(2)}|cd\rangle_{J,T}
	- \langle ab|G(\omega')|cd\rangle_{J,T}\right]^2\right\}^{1/2},
						\label{del}
\end{equation}
where the summation runs over all the $N$=332 two-body matrix elements
for the SM space consisting of the first {\em three} major shells.
Note that our full model space contains the first four major shells
but we are less concerned about the matrix elements involving the
highest shell, so we omit them from the definition of the rms
deviation.

For a fixed starting energy $\omega'$, $\delta(\omega')$
is plotted in Fig.2 as a solid curve.
{}From the figure, one sees that the smallest rms deviation of
about 0.42 MeV is obtained when the starting energy $\omega'$=75 MeV.

In Fig.2 we show as a straight dashed line, the rms
deviation when the starting energy
$\omega'$ is chosen according to the prescription in Eq.(\ref{wp})
with $\Delta$=-21 MeV.
With this prescription of the starting energy,
the G-matrix approximates $v_{\rm eff}^{(2)}$
rather well since the rms deviation is only 0.13 MeV.

\section{Applications to $^6\mbox{Li}$}

We now use the ``exact'' and ``approximate'' effective interactions
$v_{\rm eff}^{(2)}$ to perform SM calculations for $^6\mbox{Li}$.
The SM effective Hamiltonian is written as
\begin{equation}
H_{\rm SM} = \left(\sum_{i=1} t_i -T_{\rm CM}\right)
	+ \sum_{i<j}^A v_{\rm eff}^{(2)}(ij).	\label{hsm}
\end{equation}
The contribution of the center-of-mass spurious motion is removed
by adding $\lambda (H_{\rm cm}-\frac{3}{2}\hbar\Omega)$ (with $\lambda \gg1$)
to the above Hamiltonian. This is a
feature available with the {\sc oxbash} SM code \cite{oxbash}.

When one compares Eq.(\ref{ha}) and the above equation, Eq.(\ref{hsm}),
one sees that
$v_{\rm eff}^{(2)}(ij)$ is in the position of $v_{ij}$. Namely, we are
replacing the free NN potential by the effective two-body interaction.
Here we wish to point out that the SP potential ($u^{\rm HO}$)
was used only to determine the intermediate-energy spectrum in calculating
$v_{\rm eff}^{(2)}(ij)$ from $v_{ij}$ and the SP wavefunctions
of the basis space. In principle, one expects some contributions from
higher-order SP
insertions. We have not calculated them in the present investigations.
In Ref.\cite{spi}, it has been shown that higher-order
SP insertions have a negligible effect in large no-core space
SM calculations.

\begin{table}
\begin{small}
\caption{The calculated and experimental low-lying energy
spectrum for $^6\mbox{Li}$ using the ``exact'' and ``approximate''
effective interactions as discussed in the text. The results obtained
using $G^{\rm HO}(\omega')$ at a constant starting energy
($\omega'=0.0$MeV and $\omega'=75.0$MeV) are also listed.
For the ground state, the absolute
energy is given. For the excited states, the excitation energies are given.
All energies are in units of MeV.
Since we have not included the Coulomb interaction,
the experimental ground-state energy shown in the table is Coulomb corrected:
$-31.996 - E_{\rm Coulomb} = -33.996$, where
$E_{\rm Coulomb} = 2.0$ MeV is obtained from a HF calculation with the
Skyrme 3 interaction.}

\begin{center}
\begin{tabular}{c|cccc|c}\hline\hline
$J_{n}^{\pi}(T)$ & $G(\omega'=0)$ & $G(\omega'=75)$
	& ``Approx.'' & ``Exact'' & Experiment \\ \hline
$1^+_1(0)$ &-21.497&-48.386&-36.854&-35.655&-33.996\\
$3^+_1(0)$ &  2.481&  2.200&  1.916&  2.054&  2.186\\
$0^+_1(1)$ &  2.544&  5.246&  5.168&  4.932&  3.563\\
$2^+_1(0)$ &  4.955&  6.638&  6.161&  6.306&  4.31 \\
$2^+_1(1)$ &  5.660&  8.472&  8.406&  8.125&  5.37 \\
$1^+_2(0)$ &  7.514& 10.057&  9.438&  9.336&  5.65 \\
$1^+_1(1)$ & 11.295& 15.994& 15.719& 15.372&  (N/A) \\ \hline\hline
\end{tabular}
\end{center}
\end{small}
\end{table}

In Table 2, we show the results for the low-lying
energy spectrum of $^6\mbox{Li}$.
The calculations are performed in the same model space for which the
G-matrices and the effective interaction are calculated. But we only
allow up to $4\hbar\Omega$ excitations from the lowest-energy
configuration [$(0s)^4(0p)^2$].
With the HO SP basis that we used ($\hbar\Omega$=18 MeV),
$v^{(2)}_{\rm eff}$ overbinds the ground state by about 1.66 MeV,
as shown in the column labelled ``Exact''.
It should be pointed out that this result depends
quite sensitively on the HO parameter $\hbar\Omega$.
Obviously, this is related to the approximation in Eq.(\ref{appr}),
whose quality depends on $\hbar\Omega$.
Anyway, we find that when a HO basis with a smaller
$\hbar\Omega$ is used, the resulting two-body
effective interaction tends to overbind $^6\mbox{Li}$ by
an even larger amount.
We notice that this is also a feature of the results obtained by
Poppelier and Brussaard in Ref.\cite{pb}
(see Fig.7 in this reference), although in that work, the effective
interaction has some residual dependence on the starting energy.

The calculated excitation energies shown
in Table 2 for $v^{(2)}_{\rm eff}$ tend to be higher than the experimental
results but the experimental level sequence is more or less reproduced.
In Ref.\cite{pb}, the excitation energies are even higher.

It is not clear to us why the effective interactions obtained
through the Lee-Suzuki procedure
from the G-matrices using an HO SP basis with $\hbar\Omega$ smaller than
18 MeV tend to overbind
$^6\mbox{Li}$. It is quite likely that for light nuclei, the approximation
made in Eq.(\ref{appr}) of replacing the mean field generated by the
spectators by a shifted HO potential with $\hbar\Omega<18$MeV
requires significant corrections such as effective three-body forces.
Further investigations on this are necessary.

Shown in Table 2 under column ``Approx.''
are the results of the SM calculation
using the G-matrix (instead of $v^{(2)}_{\rm eff}$)
calculated at the starting
energies given by Eq.(\ref{wp}) with $\Delta=-21$MeV.
This G-matrix has been demonstrated in the previous section to be
a good approximation to $v^{(2)}_{\rm eff}$ when the
individual matrix elements are compared (see Table 1 and
Fig.1). Apparently it is also a good approximation to
$v^{(2)}_{\rm eff}$ when tested by evaluating
the energy spectrum.

In contrast, the G-matrix evaluated at any constant
starting energy $\omega'$ is not a very good approximation to
$v^{(2)}_{\rm eff}$. One can see from Table 2 that the calculated
ground-state binding energy of $^6\mbox{Li}$ using G-matrix at
$\omega'=0$ is about 14 MeV smaller than the ``exact'' result
(-21.497 MeV vs -35.655 MeV). The calculated energy spectrum
is also different. Indeed this G-matrix has
an rms deviation of about 1 MeV in its TBMEs from those of
$v^{(2)}_{\rm eff}$ (see Fig.2) and should not be expected to approximate
$v^{(2)}_{\rm eff}$ well.

When the starting energy $\omega'$ is restricted to be a constant,
the G-matrix at $\omega'=75$ best approximates $v^{(2)}_{\rm eff}$
when the rms deviation $\delta(\omega')$
in the TBMEs [Eq.(\ref{del})]
is used as the criterion (Fig.2). The SM results using
$G(\omega'=75)$ are also listed in Table 2. The ground-state
energy is clearly too low compared to the ``exact'' result.

\section{Conclusions}
We have succeeded in evaluating a starting-energy-independent
effective two-body interaction $v^{(2)}_{\rm eff}$ for a large
no-core model space for the new Nijmegen potential \cite{nijm}.
Our main conclusion is that $v^{(2)}_{\rm eff}$
can be well-approximated by the G-matrix
elements evaluated at a starting energy which depends on the
energy of the initial two-particle state in the ladder diagrams.
We have seen from Table 1 and Fig.1
that for the effective-interaction TBME
$\langle ab|v^{(2)}_{\rm eff}|cd\rangle_{J,T}$
compares favorably with the corresponding TBME of $G(\omega')$
where $\omega'$ is given by
\begin{equation}
\omega' = \epsilon^{\rm HO}_a+\epsilon^{\rm HO}_b+\Delta.
\end{equation}
For $^6\mbox{Li}$ with $\hbar\Omega$=18MeV,
we found that $\Delta=-21$MeV is a good choice. Notice that
when $\omega'$ is away from the poles of $G(\omega')$, a variation of a few MeV
in $\Delta$ is not significant as the G-matrix element is a very
slowly varying function of $\omega'$ in this case.

We emphasize that the choice of $H^{(2)}$ could be very important, as it
determines the intermediate-energy spectrum in the two-nucleon
multiple scattering processes (ladder diagrams).
Different $H^{(2)}$ will lead to different $v^{(2)}_{\rm eff}$.
In this work, we have approximated the mean field generated
by the spectator particles by a shifted HO potential, which seems
to give a reasonable description for $^6\mbox{Li}$, when the
HO characteristic parameter $\hbar\Omega$=18 MeV is used.

\section*{Acknowledgments}
We thank J. J. de Swart, M. Rentmeester and J. L. Friar for
providing us with the new Nijmegen NN potentials.
We also thank the Department of Energy for partial support
during our stay at the
Institute for Nuclear Theory at the University of Washington,
where this work was initiated.
Two of us (B.R.B. and D.C.Z.) acknowledge
partial support of this work by the National Science Foundation,
Grant No. PHY91-03011. One of us
(J.P.V.) acknowledges partial support by the U.S.
Department of Energy under Grant No. DE-FG02-87ER-40371, Division
of High Energy and Nuclear Physics and partial support from the
Alexander von Humboldt Foundation.

\pagebreak

\pagebreak

\section*{Figure Captions}

\vspace{0.2in}

\noindent
{\bf Fig.1} The phenomenological, the shifted HO and the
pure HO potential wells.
The mean field ${\cal V}(r)$ generated by the $(A-2)$
spectator nucleons may be
approximated by the phenomenological well, which is further approximated
by the shifted HO potential for convenience. In the G-matrix calculation,
we have added a constant shift $2V_0$ to both the two-nucleon
Hamiltonian $H^{(2)}$ and the starting energy $\omega$, so the pure
HO potential is used along with a shifted starting
energy $\omega'=\omega+2V_0$. The same amount of shift
added to $H^{(2)}$ and $\omega$ clearly does
not have any effect on the G-matrix, because only the difference
$\omega-H^{(2)}$ enters the G-matrix equation (\ref{go}).

\vspace{0.4in}

\noindent
{\bf Fig.2} The rms deviation $\delta(\omega')$ of the two-body matrix
elements of $G$ at a fixed starting energy $\omega'$ from those
of the starting-energy independent two-body effective
interaction $v_{\rm eff}^{(2)}$, as defined by Eq(\ref{del}) (solid curve).
The horizontal dashed line is the rms deviation
when the starting energy for the G-matrix elements is given
by the state-dependent choice in Eq.(\ref{wp}) with $\Delta=-21$MeV.

\end{document}